\documentclass[conference]{IEEEtran}
\def\BibTeX{{\rm B\kern-.05em{\sc i\kern-.025em b}\kern-.08em
    T\kern-.1667em\lower.7ex\hbox{E}\kern-.125emX}}
\IEEEoverridecommandlockouts
\usepackage{algorithm,amsbsy,amsmath,amssymb,epsfig,bbm,mathrsfs,fancyhdr,fancyvrb,url,color}
\usepackage{amsthm}
\usepackage{graphicx}
\usepackage{cite}
\usepackage{mathtools}
\usepackage{algorithm}
\usepackage{algpseudocode}
\usepackage{epstopdf}
\usepackage{epsfig}
\usepackage{tabularx}
\usepackage{pgfplots}
\usepackage{subcaption}
\usepackage{bm}
\usepackage{multicol}
\usepackage{makecell}
\usepackage{afterpage}

\usepackage{caption}
\usepackage[font=small]{caption} 
\usepackage{array}

\usepackage[english]{babel}

\newcolumntype{M}[1]{>{\centering\arraybackslash}m{#1}}

\title{OTFS-NOMA System for MIMO Communication Networks with Spatial Diversity}
\author{\IEEEauthorblockN{Wafa Hedhly, Leila Musavian, and Nikolaos Thomos}
 \thanks{The authors are with the School of Computer Science and Electronic Engineering, University of Essex, Wivenhoe Park, Colchester CO4 3SQ, United Kingdom (e-mail: { \{wafa.hedhly, leila.musavian, nthomos\}@essex.ac.uk).} This work was funded by the Engineering and Physical Science Research Council, UK, [grant number EP/X012204/1].}
 }

\begin{document}

\maketitle

\begin{abstract}
    In this work, we study the use of non-orthogonal multiple access (NOMA) and orthogonal time frequency space (OTFS) modulation in a multiple-input multiple-output (MIMO) communication network where mobile users (MUs) with different mobility profiles are grouped into clusters. 
    We consider a downlink scenario where a base station (BS) communicates with multiple users that have diverse mobility profiles. High-mobility (HM) users' signals are placed in the delay-Doppler (DD) domain using OTFS modulation in order to transform their time-varying channel into a sparse static channel, while low-mobility (LM) users signals are placed in the time-frequency (TF) domain.
    Precoding is adopted at the BS to direct focused beams towards each cluster of users. Moreover, NOMA spectrum sharing is used in each cluster to allow the coexistence of a single HM user and multiple LM users within the same resource block. LM users access disjoint subchannels to ensure their orthogonality.
    All users within the same cluster first detect the HM user's signal. Afterward, LM users suppress the interference from the HM user and detect their own signals. Closed-form expressions of the detection signal-to-noise ratios (SNRs) are derived. The numerical results showed that the performance of the proposed system highly depends on the number of LM users, the number of clusters and the power allocation factors between HM and LM users.
  
\end{abstract}

\section{Introduction}

Current and future wireless communication networks are characterized by heterogeneity on many fronts, incorporating diverse technologies and protocols and serving users with different performance requirements, channel conditions, and mobility profiles.
One way of catering to this diversity is to implement NOMA, a multiple-access paradigm that promotes user fairness and increases spectral efficiency by simultaneously sharing the resource block between users that are distinguished by their channel conditions. On this account, multiple users can be accommodated in the same resource block while mitigating the resulting interference through successive interference cancellation at the receiver \cite{islam2018resource}. On the other hand, the high mobility of users results in severe Doppler conditions and doubly dispersive channels, where the orthogonal frequency division multiplexing (OFDM) modulation fails to achieve acceptable performance \cite{raviteja2018interference}. Along this line, OTFS lately emerged as a new modulation technique where information symbols are embedded in the DD domain, yielding a sparse representation of the time-varying wireless channel.  
Motivated by these compelling attributes, we aim to study the promise of the parallel implementation of NOMA and OTFS modulation.

Several research works have investigated the potential of OTFS-NOMA in improving the performance of communication networks with heterogeneous mobility profiles \cite{ding2019otfs,ding2019robust,zhou2022active,zhou2021joint}.
In \cite{ding2019otfs}, an OTFS-NOMA transmission protocol is proposed with different mobility profiles, where a single HM user is served in the DD domain and modulated using OTFS, and LM users are modulated using OFDM. 
In \cite{ding2019robust}, a design of MIMO beamforming is proposed for a downlink OTFS-NOMA scenario where a HM user and multiple LM users are served by a BS equipped with multiple antennas. The proposed setups in \cite{ding2019otfs} and \cite{ding2019robust} are constrained by the presence of one HM user.
An uplink communication scenario was adopted in \cite{ge2021otfs} where NOMA access is implemented in the DD domain in order to accommodate stationary and mobile users using OTFS modulation for all users. However, OTFS modulation is suitable for users with high mobility profiles and high Doppler spreads.
Differently, the work in \cite{zhou2021joint} considered using OTFS modulation at all terminals in a NOMA setup for low earth orbit (LEO) constellation-enabled IoT. Similarly, in \cite{zhou2022active}, OTFS-NOMA was implemented in LEO constellations with massive MIMO IoT terminals and a successive active user identification was proposed. 
The above methods highlighted the potential of OTFS-NOMA systems, but to the best of our knowledge, most studies consider system setups with particular conditions, such as the use of single antennas or limiting the number of HM users to one. Hence, the derived conclusions cannot be generalized. Motivated by the aforementioned lack of systematic study of OTFS-NOMA in heterogeneous networks, 
 the objective of this work is to leverage the merits of OTFS-NOMA systems and MIMO precoding to convey reliable communication to multiple users with heterogeneous mobility profiles.

Thereby, we consider implementing MIMO precoding in the OTFS-NOMA system to achieve spatial diversity and serve multiple HM and LM users with the required performance.
We propose a suitable system configuration where a BS communicates with multiple HM and LM users grouped in distinct clusters. The BS performs spatial MIMO precoding in order to direct focused beams towards each cluster, where HM and LM users access the resources using NOMA paradigm. LM users access the spectrum using disjoint subchannels to ensure orthogonality among each other and, thus, fairness among users.
 Differently, HM users' signals are modulated using OTFS. All users within the same cluster detect the HM user's signal in the DD domain and cancel the resulting interference from their signals. We analyze the proposed framework and derive closed-form expressions of the signal-to-noise ratios (SNRs) of user detection. 
To thoroughly understand the performance of our system, we evaluate it following outage probability analysis examining different numbers of users per cluster, different numbers of clusters and various power allocation factors.

\section{System Model}
 \begin{figure}[!h]
	\centering 
	\captionsetup{justification=centering,margin=1cm}
	\includegraphics[trim={2.5cm 0 2.5cm 1cm},clip,width=3in]{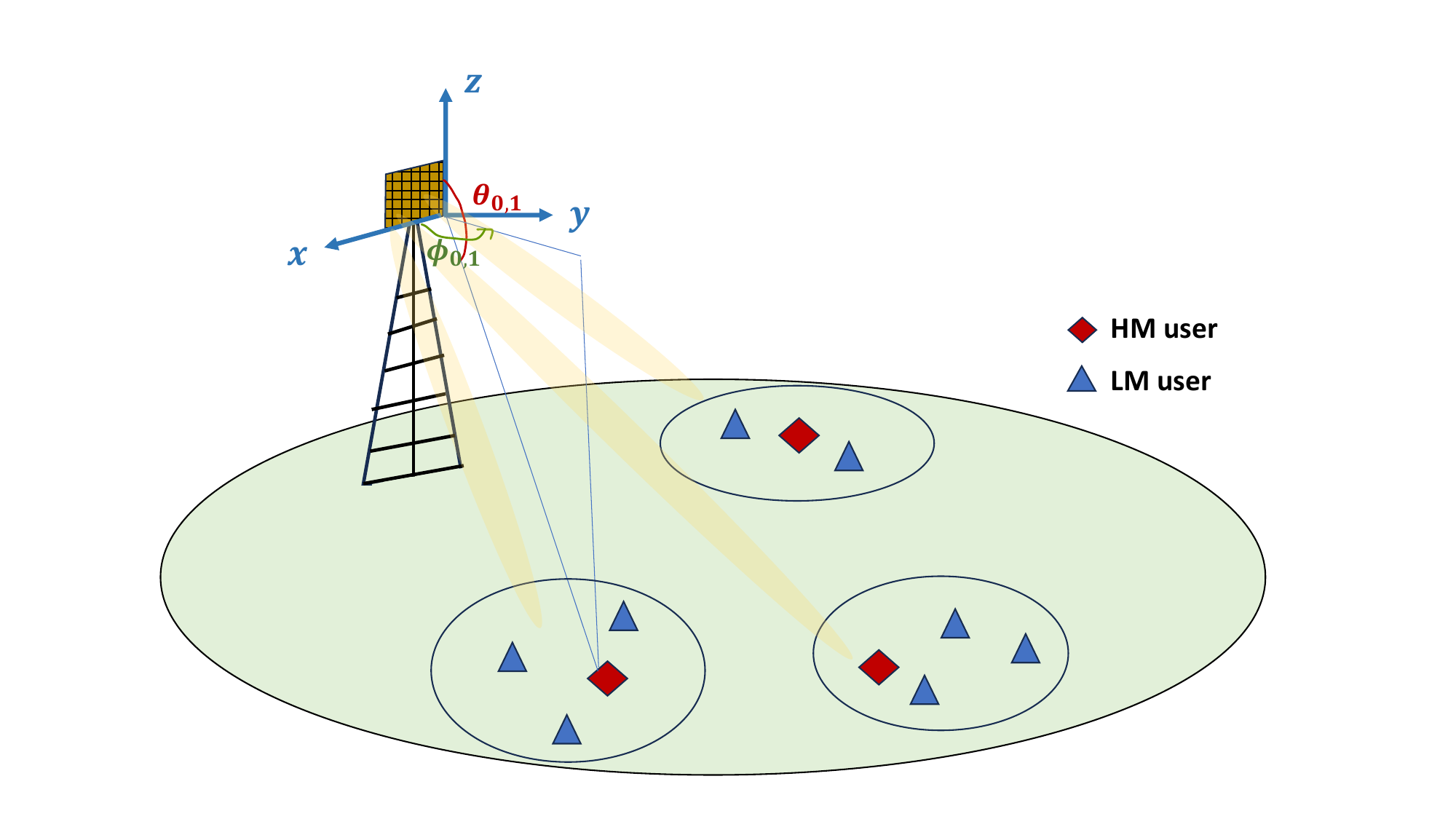}
	\caption{Proposed system setup.}
	\label{network}
\end{figure}

In this paper, we consider a communication network as depicted in Fig. \ref{network}, where a BS communicates with multiple MUs located within the cell. OTFS-NOMA is implemented in a downlink (DL) scenario.
The BS is equipped with a $W \times W$ square uniform planar array (UPA) with $A = W^2$ antenna elements, while the users are equipped with a single antenna. 
The MUs have different mobility profiles, i.e., low or high mobility. We assume multiple HM users exist in the cell and can be distinguished based on their angle-of-departures (AoDs). 
Moreover, we assume multiple LM users are distributed around each HM user. The HM user and LM users in each cluster share the spectrum using NOMA principle.
The users are grouped into $Q$ clusters based on their location. Each cluster $q$ comprises $U_q +1, 1 \le q \le Q$ users where one user ($u = 0$) is considered a HM user and the rest $U_q$ users are LM. We consider that the maximum number of LM users is the available subchannels, i.e., $ U_q \le M, \forall \; 1 \le q \le Q$, where $M$ is the number of subchannels.
The BS transmits focused beams towards each cluster of users. 
The precoding at the BS is implemented based on the location of the HM user of each cluster. Each user $u$ of cluster $q$ is located at the elevation angle $\theta_{u,q}$ and azimuth angle $\phi_{u,q}$ in the spherical coordinate system as depicted in Fig. \ref{network}.

The HM user receives OTFS-modulated signals in the DD domain, while LM users' signals are placed in the TF domain (the two domains are non-orthogonal). 
The BS implements NOMA access technique and superimposes the HM user signal modulated in the DD domain with the LM users' signals modulated in the TF domain. The HM user accesses the whole resource block while LM users occupy disjoint subchannels similarly to \cite{ding2019otfs}. 
This access technique ensures that LM users do not have frequency-selective channels and are orthogonal amongst each other. 
Each MU shares its own channel station information (CSI) with the BS.
All users within the same cluster first detect the HM user's signal of the same cluster. Then, the LM users subtract the HM user's signal from their received signal.
Last, each LM user detects its own signal.

Let us denote the transmit SNR at the BS as $\rho_{\mathrm{T}}$. We assume that the BS transmits equal power towards all LM users within the same cluster, while each cluster receives a total power based on the location of its HM user with respect to the BS.  

\subsection{Transmitted Signals}
The BS transmits an OTFS frame to each HM user with $NM$ symbols in the DD domain, where $N$ is the number of Doppler bins and $M$ is the number of delay bins drawn from a modulation alphabet. Each transmitted packet has a duration of $NT$ and occupies a bandwidth of $M \Delta f$, where $T$ is the sampling time interval and $\Delta f = 1/T$ is the spacing between subcarriers.
The DD signal is converted to the TF domain using the inverse symplectic finite Fourier transform (ISFFT) as follows,
\begin{equation}
    s_{0,q}^{\mathrm{TF}}[n,m] = \frac{1}{NM} \sum_{k = 0}^{N-1} \sum_{l = 0}^{M-1} s_{0,q}^{\mathrm{DD}}[k,l] e^{j2\pi \left( \frac{kn}{N}-\frac{ml}{M}  \right)},
\end{equation}
where $s_{0,q}^{\mathrm{DD}}[k,l]$ is the baseband transmitted binary phase-shift keying symbol at the $k^{\mathrm{th}}$ Doppler and $l^\mathrm{th}$ delay bin for $1 \le n \le N$ and $1 \le m \le M$.
For all $0 \le n \le N-1$, the TF signal of LM user $u$ of cluster $q$ is expressed as,
\begin{align}\label{suq}
    \mathbf{s}_{u,q}^{\mathrm{TF}} [n,m] = 
    \begin{cases}
        & \mathbf{s}_{u,q}^\mathrm{TF}[n, u-1], \quad \mathrm{if} \; m= u - 1, \\
        & 0, \quad \mathrm{otherwise}.
    \end{cases}
\end{align}
 The users in each cluster access the spectrum using NOMA principle. Therefore, the transmitted signal to cluster $q$ is expressed in the DD domain as,
\begin{equation}
    {s}_q[k,l] = \sum_{u = 0}^{U_q} {s}_{u,q}^{\mathrm{DD}}[k,l],
\end{equation}
where ${s}_{u,q}^{\mathrm{DD}}$ is the information bearing signal for user $u$ in cluster $q$ expressed in the DD domain.
The signal transmitted from each antenna element $(i,i'),0 \le i,i' \le W-1$, is~expressed~as,
\begin{equation}\label{xa}
    {x}_{i,i'}[k,l] = \sum_{q = 1}^Q p_{i,i',q} {s}_q[k,l],
\end{equation}
where $p_{i,i',q}$ is the precoding coefficient, $0 \le i,i' \le W-1, 1 \le q \le Q$.
The DD channel impulse response for user $u$ in cluster $q$ over $P_{u,q}$ channel paths is expressed as follows\footnote{Vectors are defined in boldface letters.},
\begin{equation}
    \mathbf{h}_{u,q}(\tau,\nu) = \sum_{p = 1}^{P_{u,q}} h_{p,u}^q \mathbf{v} \left(\theta_{p,u}^q, \phi_{p,u}^q \right) \delta (\tau - \tau_{p,u}^q) \delta (\nu - \nu_{p,u}^q),
\end{equation}
 where $h_{p,u}^q$, $\tau_{p,u}^q$ and $\nu_{p,u}^q$ are the channel coefficient, delay and Doppler of the $p^{\mathrm{th}}$ path, respectively and $\mathbf{v}$ is the steering vector depending on the elevation angle $\theta_{p,u}^q$ and azimuth angle $\phi_{p,u}^q$.
 The UPA is placed on the $(x,z)$ plane of the cartesian coordinates. Therefore, the $(i,i')^\mathrm{th}$ element of the vector $\mathrm{v}$, where $0 \le i,i' \le W-1$, is expressed as \cite{zaharis2022improved}, 
\begin{equation}
    v_{ii'}(\theta,\phi) = \exp{ \left( j2\pi \frac{d}{\lambda} \left(i\sin(\theta)\cos(\phi) +i' \cos(\theta) \right) \right)},
\end{equation}
where $d$ is the separation distance between the UPA antenna elements assumed to be equal in the two dimensions, and $\lambda$ is the wavelength.
In the case of LM users, the channel is not subject to Doppler shifts since they have low mobility. Therefore, their Doppler shifts are $\nu_{p,u}^q = 0, \forall \; 1 \le u \le U_q$ and $1 \le q \le Q$.

\subsection{Received Signals}
Assuming bi-orthogonality between the transmit and receive waveforms, the received signal at the HM user of cluster $q$ from antenna element $(i,i')$, $0 \le i,i' \le W-1$ is expressed in the DD domain as \cite{raviteja2018interference},
 \begin{align}\label{y0}
    y_{0,q}^{i,i'}[k,l] &=  \sum_{p = 1}^{P_{0,q}} h_{p,0}^q e^{-j2\pi \tau_{p,0}^q \nu_{p,0}^q} v_{ii'} \left(\theta_{p,0}^q,\phi_{p,0}^q \right) \times \nonumber \\ & x_{i,i'}[\left( k-k_{p,0}^q  \right)_N,\left( l-l_{p,0}^q  \right)_M] + w_{0,q}[k,l],
 \end{align}
where $w_{0,q}$ denotes the complex Gaussian additive noise with zero mean and variance $\sigma_{\mathrm{w}}^2$, $x_{i,i'}^\mathrm{DD}$ is the transmitted signal, $k_{p,0}^q$ and $l_{p,0}^q$ are the indices for the Doppler and delay taps of path $p$, respectively\footnote{Herein, we assume the OTFS grid resolution is high enough to neglect fractional Doppler. $(x)_N$ denotes $x$ modulo $N$.}. 
The Doppler and delay taps are expressed as,
\begin{equation}\label{taps}
    \nu_{p,0}^q = \frac{k_{p,0}^q}{NT} \quad \mathrm{and} \quad \tau_{p,0}^q = \frac{l_{p,0}^q}{M \Delta f}.
\end{equation}
Similarly to \cite{singh2022low}, we assume that the path gains $h_{p,u}^q, 1 \le p \le P_{u,q} $ are 
independent and identically distributed (i.i.d) $\mathcal{C}(0,\sigma_{\mathrm{p}}^2)$ with a uniform scattering profile, i.e., $\sigma_{\mathrm{p}}^2 = 1/P_{u,q}$. We also assume that the path delay and Doppler taps have the same locations between all transmit antennas and the user receiver. 

The received signal at user $u$ in cluster $q$ is expressed as,
\begin{equation}
    y_{u,q} = \sum_{i = 1}^W \sum_{i' = 1}^W y_{u,q}^{i,i'},
\end{equation}
where $y_{u,q}^{i,i'}$ is the signal received by user $u$ from antenna element $(i,i')$. 
To simplify the notations, hereafter, we substitute the $(i,i')$ of each antenna element by the single index $a$, where $1 \le a \le A$. 
As a result, the received signal vector at each user $u$ in cluster $q$ is expressed as \cite{pandey2021low},
\begin{equation}
    \mathbf{y}_{u,q} = \sum_{a = 1}^A \mathbf{H}_{a,u,q} \mathbf{x}_{a} + \mathbf{w}_{u,q},
\end{equation}
where $\mathbf{H}_{a,u,q} \in \mathcal{C}^{N\!M \times N\!M}$ is a block circulant channel matrix between user $u$ and antenna $a$ ($M$ circulant blocks of $N \times N$ circulant matrices) and $\mathbf{w}_{u,q} \sim \mathcal{CN}(\mathbf{0}, {\sigma}_{\mathrm{w}}^2 \mathbf{I}_{N \! M})$ is the i.i.d additive noise vector in the DD domain.
The signal vector $\mathbf{x}_{a}$ is constructed from $x_{a}$ expressed in \eqref{xa}  while satisfying the following condition: the $(k+Nl)^{\mathrm{th}}$ element of $\mathbf{x}_{a}$ is equal to $x_a [k,l]$. The vectors $\mathbf{y}_{u,q}$ and $\mathbf{w}_{u,q}$ are similarly constructed.

To simplify the notations, we remove the cluster index $q$ and consider the index $u = 0$ for the HM user. Without loss of generality, we consider the users of cluster 1.
Therefore, the received signal in the DD domain at user $u$ of cluster 1 is expressed as,
\begin{equation}\label{yu}
    \mathbf{y}_{u} = \sum_{a = 1}^A \mathbf{H}_{a,u} \mathbf{x}_{a} + \mathbf{w}_{u}, 0 \le u \le U_1.
\end{equation}

The received signal \eqref{yu} at user $u$ can be re-written as,
\begin{align}
    \mathbf{y}_u &= \sum_{a = 1}^A p_{a,1} \mathbf{H}_{a,u}  \mathbf{s}_{0,1} + \sum_{a = 1}^A p_{a,1} \mathbf{H}_{a,u}  
 \sum_{i = 1}^{U_1} \mathbf{s}_{i,1} \nonumber \\
 & + \sum_{a = 1}^A \mathbf{H}_{a,u}  \sum_{q = 2}^Q p_{a,q} \mathbf{s}_q + \mathbf{w}_{u},
\end{align}
where $ p_{a,1}$ is the precoder for cluster 1.

\section{Users Detection}
\subsection{High-Mobility User Detection}
As explained earlier, all users within each cluster first detect the HM user's signal. The system utilizes zero-forcing (ZF) to detect the LM user at each user of the same cluster. 
Since the channel matrix $ \mathbf{H}_{a,u}$ is a circulant matrix with $M$ circulant blocks, we express the OTFS channel matrix 
between antenna element $a$ and  user $u$ using the following decomposition \cite{singh2022ber},
\begin{equation}\label{diag}
    \mathbf{H}_{a,u} = \mathbf{\Psi}^\mathrm{H} \mathbf{D}_{a,u} \mathbf{\Psi},
\end{equation}
where $\mathbf{\Psi} = \mathbf{F}_N \otimes \mathbf{F}_M$ is an $N\!M \times N\!M$ matrix, $\mathbf{F}_N$ and $\mathbf{F}_M$ are, respectively, $N \times N$ and $M \times M$ discrete Fourier transform (DFT) matrices\footnote{$\mathbf{A} \otimes \mathbf{B}$ denotes the Kronecker product of the matrices $\mathbf{A}$ and $\mathbf{B}$. The superscript H denotes the Hermitian transform.}.
The diagonal matrix $\mathbf{D}_{a,u}$ is composed of the eigenvalues of $\mathbf{H}_{a,u}$ and can be further expressed as,
\begin{equation}
    \mathbf{D}_{a,u} = \sum_{i = 0}^{M-1} \mathbf{O}^i \otimes \mathbf{D}_{a,u,i},
\end{equation}
where $\mathbf{O} = \mathrm{diag} \left[1, e^{j2\pi/M}, \ldots, e^{j2\pi(M-1)/M}   \right]$ and $\mathbf{D}_{a,u,i}$ is a diagonal matrix consisting of the eigenvalues of the $i^\mathrm{th}$ circulant block of $\mathbf{H}_{a,u}$. This decomposition significantly reduces the computation complexity of the system.
We consider the following notation,
\begin{equation}\label{hu}
    \mathbf{H}_u = \sum_{a = 1}^A p_{a,1} \mathbf{H}_{a,u}.
\end{equation}
Using the diagonalization of $\mathbf{H}_{a,u}$ as shown in \eqref{diag}, the matrix $\mathbf{H}_u$ can be re-written as,
\begin{equation}
    \mathbf{H}_u = \mathbf{\Psi}^\mathrm{H} \mathbf{D}_u \mathbf{\Psi},
\end{equation}
where the diagonal matrix $\mathbf{D}_u$ is expressed as,
\begin{equation}
    \mathbf{D}_u =  \sum_{a = 1}^A p_{a,1} \mathbf{D}_{a,u}.
\end{equation}
Therefore, the equalization matrix to detect $\mathbf{s}_{0,1}$ is~as~follows,
\begin{equation}
    \mathbf{G}_u =  \left( \mathbf{H}_u^\mathrm{H} \mathbf{H}_u  \right)^{-1} \mathbf{H}_u^\mathrm{H}.
\end{equation}
Since, $\mathbf{\Psi} \mathbf{\Psi}^\mathrm{H} = \mathbf{\Psi}^\mathrm{H}\mathbf{\Psi}  = \mathbf{I}_{MN}$, the equalization matrix $\mathbf{G}_u$ can be re-expressed as,
\begin{equation}
    \mathbf{G}_u = \mathbf{\Psi}^\mathrm{H} \left( \mathbf{D}_u^{\mathrm{H}} \mathbf{D}_u \right)^{-1} \mathbf{D}_u^{\mathrm{H}} \mathbf{\Psi}.
\end{equation}
If we denote by $\lambda_{a,u,i}, 1 \le i \le NM$ the eigenvalues of $\mathbf{H}_{a,u}$, then the matrix $\mathbf{G}_u$ can equivalently be written as,
\begin{equation}
    \mathbf{G}_u = \mathbf{\Psi}^{\mathrm{H}} \mathbf{\Delta}_u \mathbf{\Psi},
\end{equation}
where $\mathbf{\Delta}_u$ is a diagonal matrix consisting of elements,
\begin{equation}
    \delta_{u,i} = \frac{\sum_{a = 1}^A p_{a,1}^* \lambda_{a,u,i}^*}{\left |\sum_{a = 1}^A p_{a,1} \lambda_{a,u,i} \right|^2 }, 1 \le i \le NM.
\end{equation}
The HM user detects its own signal while considering the LM users' signals as noise. LM users detect and subtract the HM user signal from the received signal. 
The proposed equalizer is applied to the received signal at user~$u $~as~follows,
\begin{equation}
   \mathbf{G}_u  \mathbf{y}_u = \mathbf{{s}}_{0,1}  + \mathbf{\kappa}_u, \quad 0 \le u \le U_1,
\end{equation}
where the interference-plus-noise term is written as,
\begin{equation}
    \mathbf{\kappa}_u =  \sum_{i = 1}^{U_1} \mathbf{{s}}_{i,1} + \mathbf{G}_u \left( \sum_{a = 0}^A \mathbf{H}_{a,u}  \sum_{q = 2}^Q p_{a,q} \mathbf{s}_q + \mathbf{w}_{u} \right).
\end{equation}
The interference consists of intra-cluster interference from LM users' signals and inter-cluster interference from users in other clusters.
The interference-plus-noise term can be equivalently~written~as,
\begin{equation}\label{Iu}
    \mathbf{\kappa}_u = \sum_{i\neq u}^{U_1} \mathbf{{s}}_{i,1} + \sum_{q=2}^Q \mathbf{\Omega}_{q,u} \mathbf{s}_q + \mathbf{z}_u,
\end{equation}
where the noise term is given by $\mathbf{z}_u =  \mathbf{G}_u \mathbf{w}_u$ and the matrix $\mathbf{\Omega}_{q,u}$ is given by,
\begin{equation}\label{omega}
    \mathbf{\Omega}_{q,u} = \mathbf{\Psi}^\mathrm{H} \mathbf{\Delta}_u   \sum_{a = 1}^A p_{a,q} \mathbf{D}_{a,u} \mathbf{\Psi}.
\end{equation}
Therefore, the SINR of the $k^{\mathrm{th}}$ symbol of the HM user's transmit vector at user $u$ is expressed as,
\begin{equation}\label{snr0}
    \gamma_u^0 = \frac{\rho_0}{\sum_{i = 1}^{U_1} \rho_u +  \sum_{q = 2}^Q  \rho_{c,q} \omega_{q,u} + \eta_u }, 
\end{equation}
where $\omega_{q,u}$ and $\eta_u$ are written,
\begin{align}
    \omega_{q,u} &= \frac{1}{NM} \sum_{i = 1}^{NM} \left| \frac{  \sum_{a = 1}^A p_{a,q} \lambda_{a,u,i} }{\sum_{a = 1}^A p_{a,1} \lambda_{a,u,i} } \right|^2 \nonumber \\
    \eta_u &= \frac{1}{NM} \sum_{i = 1}^{NM}  \frac{  1 }{\left| \sum_{a = 1}^A p_{a,1} \lambda_{a,u,i}\right|^2 },
\end{align}
and $\rho_u$ is the transmit SNR of the $u^{\mathrm{th}}$ user of cluster 1 and $\rho_{c,q}$ is the transmit SNR to cluster $q$. Therefore, $\rho_0 + \sum_{i = 1}^{U_1} \rho_u = \rho_{c,1}$. 
See appendix \ref{snr} for the proof. 

\subsection{Low-Mobility User Detection}
After successfully decoding and subtracting the HM user's signal, each LM user $u, 1 \le u \le U_1$ observes the following received signal in the TF domain,
\begin{align}\label{yu1}
    y_u[n,m] &= \sum_{a = 1}^A p_{a,1} {H}_{a,u}[n,m] \sum_{i = 1}^{U_1} s_{i,1}[n,m] \nonumber \\
    &+ \sum_{a = 1}^A  {H}_{a,u}[n,m] \sum_{q = 2}^Q p_{a,q} s_{q}[n,m] + w_u[n,m],
\end{align}
where $H_{a,u}[n,m]$ is the channel gain between antenna $a$ and user $u$ and $w_u[n,m]$ is the noise in the TF domain. The samples are conducted at instant $nT$ and channel $m\Delta f$.
The TF domain channel gain can be expressed as \cite{singh2022low},
\begin{equation}\label{hnm}
    H_{a,u}[n,m] = \int_\tau \int_\nu h_{a,u}(\tau,\nu) e^{j2\pi \nu nT} e^{-j2\pi (\nu+m\Delta f)\tau} d\tau d\nu,
\end{equation}
where $ h_{a,u}(\tau,\nu)$ is the channel impulse response between antenna $a$ and user $u$.
Considering the absence of Doppler shifts in low-mobility scenarios, the channel gain in \eqref{hnm} can be simplified to a time-independent expression as follows,
\begin{equation}
    H_{i,i',u}[n,m] = \sum_{p = 1}^{P_u} h_{p,u} e^{j2\pi m \Delta f \tau_{p,u}} v_{ii'}\left(\theta_{p,u},\phi_{p,u} \right).
\end{equation}
Since $\tau_{p,u} = l_{p,u}/(M \Delta f)$ where $l_{p,u}$ is the delay tap pf path $p$,
the channel gain can be written as,
\begin{equation}
    H_{i,i',u}[m] = \sum_{p=1}^{P_u} h_{p,u} e^{j2\pi \frac{l_{{p,u}}m}{M}} v_{ii'}\left(\theta_{p,u},\phi_{p,u} \right).
\end{equation}
Considering the transmitted signal expression in \eqref{suq} and thanks to the orthogonality between LM users within the same cluster, user $u$ disregards the signals received on subchannels $m \neq u - 1$ and the expression in \eqref{yu1} can be re-written as,
\begin{align}
   & y_u[n,u-1] = \sum_{a = 1}^A p_{a,1} {H}_{a,u}[u-1] s_{u,1}[n,u-1] \nonumber \\
    &+\sum_{q = 2}^Q  \sum_{a = 1}^A  {H}_{a,u}[u-1]  p_{a,q} 
 s_{q}[n,u-1] 
    + w_u[n,u-1].
\end{align}
Since the bandwidth of each subcarrier is considered smaller than the coherence bandwidth of the channel, one-tap equalization can be used to detect the LM user's signal \cite{ding2019otfs}. The equalized signal of LM user $u$ at time instant $n$ is~expressed~as,
\begin{align}
   & \frac{y_u[n,u-1]}{H_u[u-1]} = s_u[n,u-1] \nonumber \\
    &+ \sum_{q = 2}^Q \frac{\sum_{a = 1}^A p_{a,q} H_{a,u}[u-1]}{H_u[u-1]} s_q[n,u-1] + \frac{w_u[n,u-1]}{H_u[u-1]}.
\end{align}
 \begin{figure}[!h]
	\centering 
	\captionsetup{justification=centering,margin=0.5cm}
	\includegraphics[width=3in]{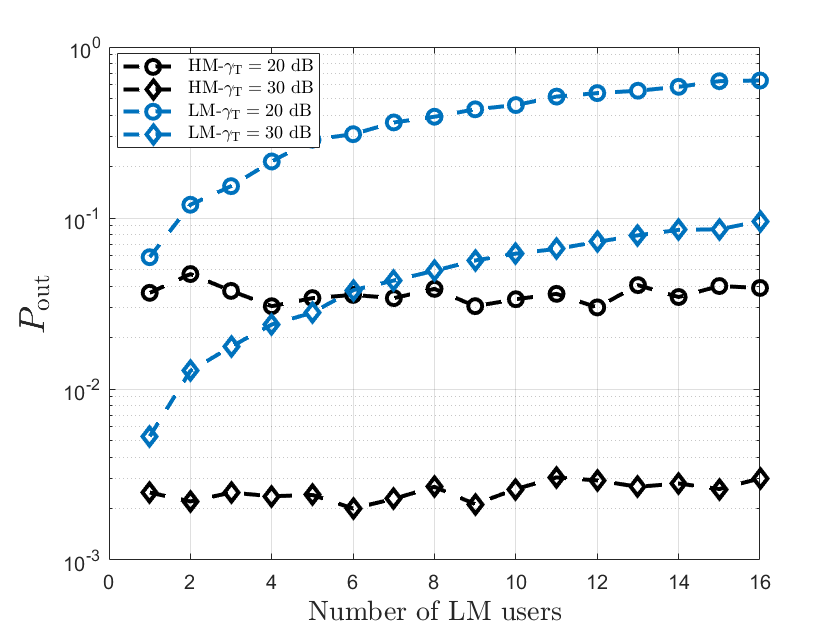}
	\caption{Outage probability of HM and LM users versus the number of LM users for different modulation schemes.}
	\label{outU}
\end{figure}
Since LM users are time-invariant, the SINR expression is the same for all time instants. Therefore, the SINR for detecting the user $u$ is given by,
\begin{equation}
    \gamma_u = \frac{\rho_u \left|{H_u[u-1]}\right|^2}{ \sum_{q = 2}^Q  \rho_{c,q} \left| {\sum_{a = 1}^A p_{a,q} H_{a,u}[u-1]}  \right| ^2 + 1}.
\end{equation}

\section{Simulation Results}
To evaluate the performance of the proposed system, we consider the following simulation parameters: $A = 64, N = 16, M = 16, \Delta f = 15 \; \mathrm{kHz}, f_{\mathrm{c}} = 60 \; GHz$. 
As in previous studies \cite{raviteja2018interference}, Doppler taps are generated according to Jake's model, considering that the maximum vehicle speed is 200 km/h.
The maximum channel delay tap is $l_{\max} = 4$. The number of paths is 4 for all users. The elevation and azimuth AoAs of paths are uniformly distributed with mean $\pi/4$ and $0$, respectively, and with variance $\pi/10$ and $\pi$, respectively. 
Unless otherwise specified, we consider the following parameters: $Q = 3$, $\rho_{\mathrm{T}} = 30 \; \mathrm{dB}$.
The SNR of the HM user of cluster 1 is $\rho_0 = \alpha \rho_{c,1}$, where $\alpha = 3/4$. Therefore, the SNR of each LM user of cluster 1 is $\rho_u = {(1-\alpha) \rho_{c,1}}/{U_1}, 1 \le u \le U_1$. 
The number of LM users in each cluster is selected uniformly at random in $[ \, 1, M] \,$. The LM users' positions are uniformly placed in a circle around the HM user of each cluster with a radius $R_{\mathrm{Q}} = 10 \; \mathrm{m}$. In the following simulation examples, the outage probability is the probability that the rate falls below the threshold $R_\mathrm{th} = 0.5 \; \mathrm{b/s/Hz}$.

 \begin{figure}[!h]
	\centering 
	\captionsetup{justification=centering,margin=0.5cm}
	\includegraphics[width=3in]{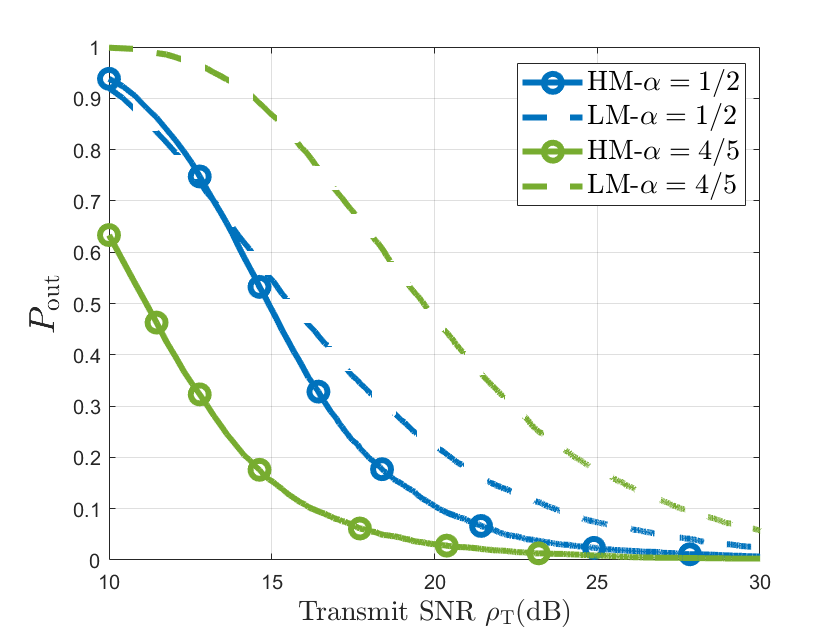}
	\caption{Outage probability of HM and LM users for different power allocation factors.}
	\label{outalpha}
\end{figure}
 \begin{figure}[!h]
	\centering 
	\captionsetup{justification=centering,margin=0.5cm}
	\includegraphics[width=3in]{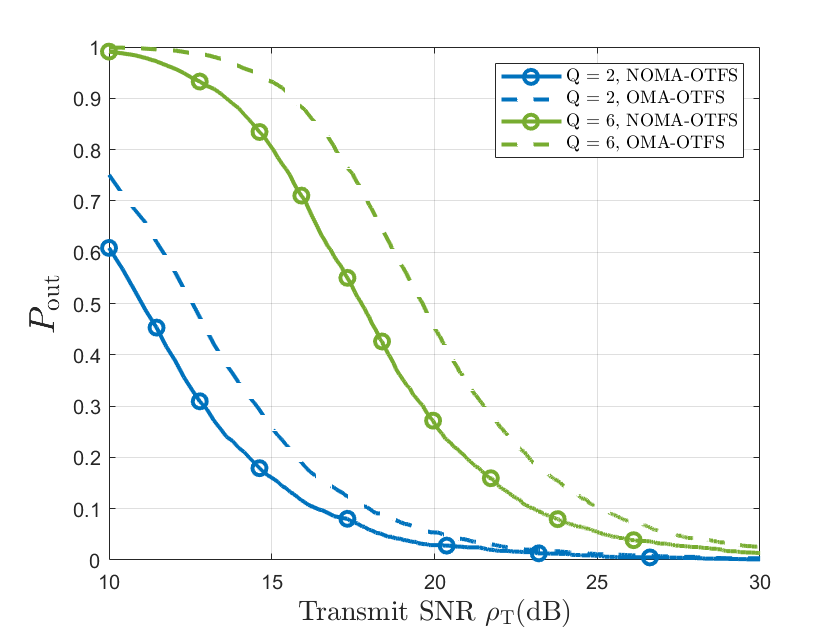}
	\caption{Benchmark comparison of the proposed OTFS-NOMA system with OMA access.}
	\label{benchmark}
\end{figure}
 
In the first simulation example, we investigate the effect of the number of LM users within each cluster on the system performance. To this end, we plot in Fig. \ref{outU} the outage probability of the HM and LM users versus the number of LM users for various transmit SNRs.
First, we observe that the outage probability of the HM user is quasi-constant as a function of the number of LM users within the cluster. This result is valid for both values of the transmit SNR. This can be justified by the power allocation scheme that is adopted by the system. The BS allocates a fixed fraction of the power to the HM user and distributes the remaining power equally among the LM users. Moreover, since each LM user has access to a single subchannel, some subchannels are unused when $U_q < M$. Hence, we note that the behavior of the HM outage probability remains constant, implying that the effect of the distribution of interference coming from LM users between subchannels does not have a significant impact on the performance of the HM user. In other words, given that the HM OTFS frame spans the whole resource block, the HM user's performance remains roughly the same whether the interference is concentrated within a narrow frequency band or spread over a larger band. 
Second, we note from Fig. \ref{outU} that increasing transmit SNR from 20 dB to 30 dB has a significant performance improvement in terms of quality-of-service (QoS) delivered to the users. Although the decrease of the outage probability is more significant at the HM user's side, LM users also benefit from the resource boost. 
Since the transmit power allocated for LM users decreases with their number, increasing the transmit SNR or tuning the allocated power to LM users helps compensate for the performance degradation and reach acceptable QoS.

In the second simulation example, we investigate the effect of the power allocation factor on the system performance. In Fig. \ref{outalpha}, the outage probability is shown versus the transmit SNR at the BS for different allocation factors. We note that the power distribution between the HM user and LM users has a significant performance impact, especially for low SNRs. In high SNR regimes, this effect becomes less significant.
Moreover, when $\alpha = 4/5$, the performance gap between the HM and LM users is much more significant than the performance gap in the case of $\alpha = 1/2$. Subsequently, increasing the transmit power of the HM user will certainly decrease its outage probability, but at the same time, it can deteriorate the performance of LM users in the process. 
As a result, it is important to strategically tune the power allocation factor in order to keep acceptable QoS~for~all~users.

In the last simulation example, we compare the proposed NOMA system configuration with an orthogonal multiple access (OMA) scheme. Herein, we adopt the time division multiple access between HM and LM users. To this end, in Fig. \ref{benchmark}, we present the outage probability of the HM user's signal detection at the HM user's side versus the total DL transmit SNR for different numbers of clusters. We can observe that the proposed OTFS-NOMA scheme outperforms its orthogonal counterpart and provides more reliable communication links for all the examined values of the transmit SNR. Also, the performance gap between the proposed NOMA setup and OMA benchmark increases with the number of clusters within the cell. This behavior suggests that the reliability of the proposed system becomes more significant and impactful in low SNR regimes.

\section{Conclusion}
In this paper, we proposed an OTFS-NOMA system where spatial diversity is adopted to convey data signals to multiple clusters of users simultaneously. Using NOMA paradigm, the BS superimposes an OTFS signal frame transmitted to the HM user and TF-modulated signals transmitted to  LM users. The LM users access disjoint subchannels. Subsequently, their signals are, by design, orthogonal to each other. The results demonstrated that the power distribution between users and the number of LM users have a remarkable impact on the outage probability of LM users. 
However, the performance of the HM user remains unchanging when the number of LM users increases. 
 Moreover, the proposed framework outperforms the OMA setup, especially with higher numbers of clusters.

\begin{appendices}
    \section{SNR Expression} \label{snr}
    The SNR of $k^{\mathrm{th}}$ symbol detection is expressed as,
    \begin{footnotesize}
    \begin{equation}
        \gamma_u^0 = \frac{\mathbb{E} \left(|\mathbf{s}_{0,1}[k]|^2 \right)}{\sum_{i = 1}^{U_1}\! \mathbb{E} \left( \left| \mathbf{s}_{i,1}[k] \right|^2 \right) \! + \!\sum_{q = 2}^Q \! \mathbb{E}   \left( \mathbf{\Omega}_{q,u} \mathbf{s}_q \mathbf{s}_q^\mathrm{H} \mathbf{\Omega}_{q,u}^\mathrm{H}  \right)_{k,k} \!+ \!\mathbb{E} \left( \mathbf{z}_u \mathbf{z}_u^\mathrm{H} \right)_{k,k} } 
    \end{equation}
    \end{footnotesize}
Taking into account that $\mathbf{w}_{u,q}\! \sim \! \mathcal{CN}(\mathbf{0}, {\sigma}_{\mathrm{w}}^2 \mathbf{I}_{N \! M})$ and $\mathbf{G}_u  \mathbf{G}_u^\mathrm{H}\! = \! \left( \mathbf{H}_u^\mathrm{H}  \mathbf{H}_u \right)^{-1} $, the noise covariance matrix is 
 $ \mathbb{E} \left( \mathbf{z}_u \mathbf{z}_u^\mathrm{H} \right) \!=\! \mathbb{E} \left( \mathbf{G}_u \mathbf{w}_u \mathbf{w}_u^\mathrm{H} \mathbf{G}_u^\mathrm{H} \right) 
    \! =\! \sigma_{\mathrm{w}}^2 \left( \mathbf{H}_u^\mathrm{H}  \mathbf{H}_u \right)^{{-1}}.$
     We assume that the transmit power for cluster $q$ is $\sum_{u = 0}^{U_q} P_{u,q} = P_q, \forall q \in \{1, \ldots, Q \}$. The vector $\mathbf{s}_q \sim \mathcal{CN}(\mathbf{0}, \sum_{0\le u \le U_q} P_{u,q} \mathbf{I}_{N \! M})$.
Thus, the covariance of the inter-cluster interference term can be re-written as,
\begin{align}
    \sum_{q = 2}^Q \mathbb{E} \left( \mathbf{\Omega}_{q,u} \mathbf{s}_q \mathbf{s}_q^\mathrm{H} \mathbf{\Omega}_{q,u}^\mathrm{H}  \right)
    &=   \sum_{q = 2}^Q P_q \mathbf{\Omega}_{q,u}  \mathbf{\Omega}_{q,u}^\mathrm{H} .
\end{align}

If we denote by $\eta_{u,i}$ the $i^{\mathrm{th}}$ eigenvalue of $\mathbf{H}_u$, then, according to \eqref{hu}, it can be expressed as $\eta_{u,i} = \sum_{a = 1}^A p_{a,1} \lambda_{a,u,i}$.
According to \cite{singh2022ber}, the term $\left(  \left( \mathbf{H}_{u}^\mathrm{H} \mathbf{H}_{u} \right)^{-1} \right)_{k,k}$ can be simplified as follows,
\begin{equation}
    \left(  \left( \mathbf{H}_{u}^\mathrm{H} \mathbf{H}_{u} \right)^{-1} \right)_{k,k} = \frac{1}{NM} \sum_{i = 1}^{NM} \frac{1}{\left| \eta_{u,i} \right|^2}.
\end{equation}
Similarly, if we denote by $\omega_{q,u,i}$ the $i^{\mathrm{th}}$ eigenvalue of $\mathbf{\Omega}_{q,u}$, then, according to \eqref{omega}, it can be expressed as,
\begin{equation}
    \omega_{q,u,i} =  \frac{\sum_{a = 1}^A p_{a,1}^* \lambda_{a,u,i}^*}{\left |\sum_{a = 1}^A p_{a,1} \lambda_{a,u,i} \right|^2 }  \sum_{a = 1}^A p_{a,q} \lambda_{a,u,i}.
\end{equation}
The $(k,k)^{\mathrm{th}}$ element of $\mathbf{\Omega}_{q,u}  \mathbf{\Omega}_{q,u}^\mathrm{H}$ can be simplified as follows,
\begin{equation}
    \left(\mathbf{\Omega}_{q,u}  \mathbf{\Omega}_{q,u}^\mathrm{H}\right)_{k,k} = \frac{1}{NM} \sum_{i = 1}^{NM} {\left| \omega_{q,u,i} \right|^2}.
\end{equation}
Normalizing the SNR by $\sigma_\mathrm{w}^2$, we get the expression in \eqref{snr0}.
\end{appendices}

\bibliographystyle{IEEEtran}
\bibliography{IEEEabrv,references}

\end{document}